\documentclass[final,5p,times,twocolumn]{elsarticle}

\usepackage{tabularx}
\usepackage{booktabs}
\usepackage{soul}
\usepackage{xcolor}
\usepackage{float}
\definecolor{lightred}{rgb}{1.0,0.6,0.6}
\sethlcolor{lightred}
\usepackage[breaklinks=true]{hyperref}

\usepackage{amssymb}
\usepackage{amsmath}
\journal{Applied Radiation and Isotopes}

\begin{document}

% \linenumbers

\begin{frontmatter}



\title{High efficiency quantification of $^{90}$Sr contamination in cow milk after a nuclear accident}


\author[HEPIA]{Q. Rogliardo} %% Author name
\author[HEPIA]{A. Kanellakopoulos}
\author[HEPIA]{H. Corcelle}
\author[HEPIA]{M. Fedel}
\author[HEPIA]{M. Zsely-Schaffter}
\author[HEPIA]{G. Triscone}
\author[HEPIA]{S. Pallada\corref{cor}}
\ead{stavroula.pallada@hesge.ch}

\cortext[cor]{Corresponding author.}

\affiliation[HEPIA]{organization={Nuclear Physics \& Chemistry Laboratories, HEPIA HES‐SO, University of Applied Sciences and Arts Western
Switzerlanda},%Department and Organization
            addressline={Rue de la Prairie 4}, 
            city={Geneva},
            postcode={1202},
            country={Switzerland}}

%% Abstract
\begin{abstract}
Monitoring $^{90}$Sr contamination in milk following a nuclear accident is critical due to its radiotoxicity and calcium-mimicking behaviour, leading to accumulation in bones and teeth. This study presents a high-efficiency protocol for quantifying $^{90}$Sr in cow milk by integrating freeze-drying, high-temperature calcination, ion exchange chromatography and  liquid scintillation spectroscopy (LSC). The method was validated using reference milk samples with 0.45~Bq/mL of $^{90}$Sr, achieving a chemical yield of 100 $\pm$ 2\%, ensuring near-complete recovery and accurate quantification. 

The minimum detectable activity (MDA) was estimated at 0.33 Bq/L under optimal conditions, demonstrating the protocol's sensitivity for low-level detection. A comparative analysis with existing methods, centrifugation-based approaches and Dowex resin techniques revealed that our protocol outperforms in both strontium recovery and organic matter elimination. Alternative methods showed lower recovery rates (68 $\pm$ 2\% for Guérin's method, 65 $\pm$6\% for Dowex resin) and suffered from procedural drawbacks, such as incomplete organic matter removal.

Applying this methodology to compare samples from certified laboratories confirmed its robustness, with liquid scintillation spectroscopy radioactivity values doubling after 14 days, consistent with secular equilibrium between $^{90}$Sr and $^{90}$Y. While the protocol is optimized for milk, future research should explore its applicability to other food matrices. The high yield, reliability, and ease of implementation, position this method as an effective tool for nuclear emergency response and routine radiological monitoring.
\end{abstract}

%% Keywords

\end{frontmatter}

\section{Introduction}
In recent years, growing concerns over nuclear safety have intensified due to potential reactor accidents, unauthorized releases in the environment, and geopolitical conflicts involving nuclear weapons. In response, multiple European countries, alongside national and international agencies, have been developing environmental monitoring protocols for nuclear emergencies \cite{IAEA, Duliba2023}. Many of these protocols prioritize food safety, with a particular emphasis on milk, given its widespread consumption and its essential role in child nutrition. 

Radiocontaminated milk containing $^{90}$Sr presents significant challenges due to its potential to propagate contamination throughout the food chain \cite{Michon1969}. $^{90}$Sr, a major radioisotope released during nuclear accidents, is a high-energy $\beta$\textsuperscript{-} emitter with calcium-mimicking chemical and biological properties. This mimicry raises major health concerns, as $^{90}$Sr can become incorporated into teeth and bone tissue, leading to long-term radiotoxic effects.

\begin{figure}[h]
    \centering
	\includegraphics[width=0.7\linewidth]{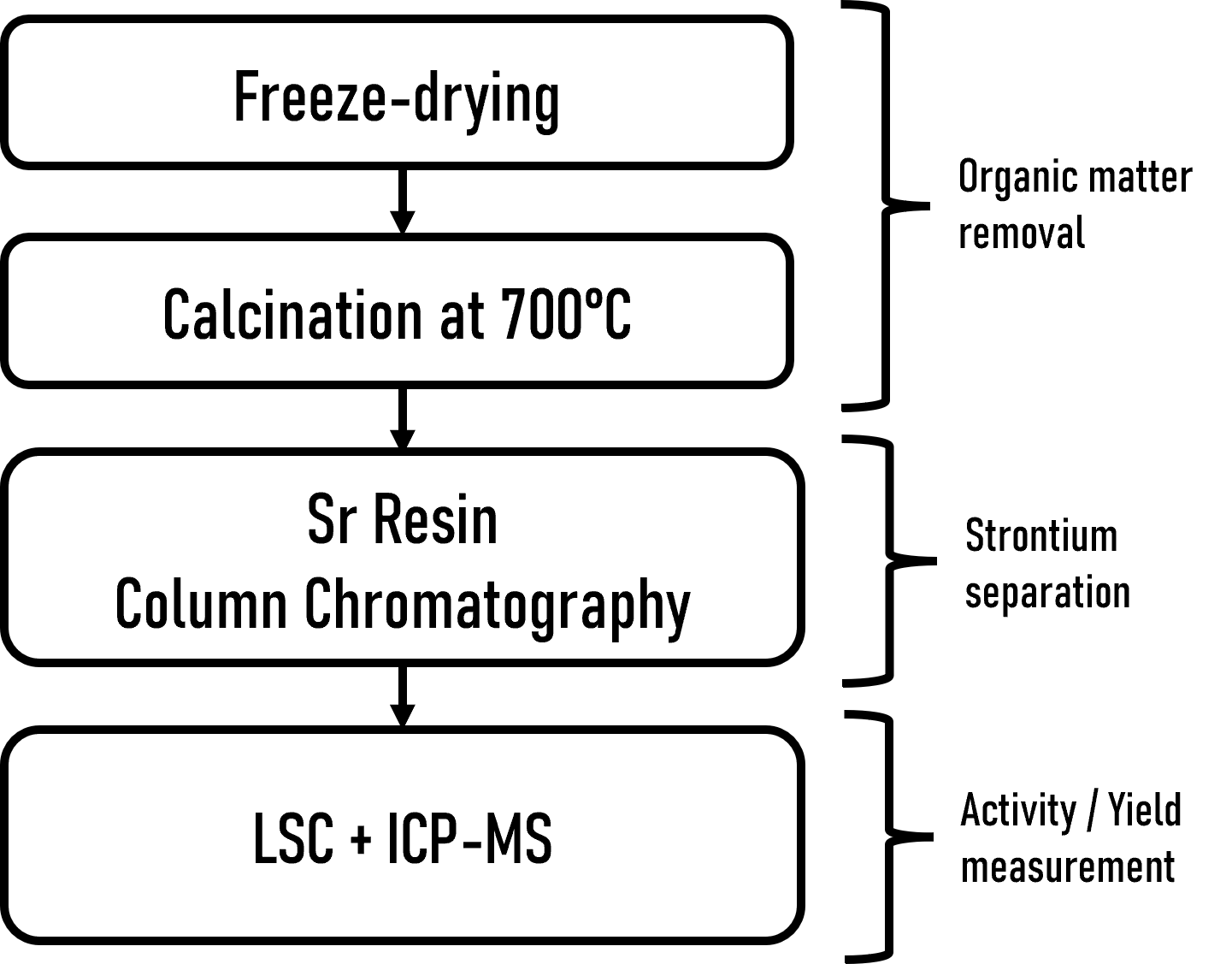}
	\caption{Flow chart of the developed protocol.}
	\label{fig:flowchart}
\end{figure}

This work presents a safe, efficient and user-friendly protocol for quantifying $^{90}$Sr radioactivity in milk, structured around three key steps: elimination of organic matter, selective separation of strontium and measurement of the extracted radioactivity. The protocol begins with freeze-drying, followed by calcination at 700°C to remove organic content. Strontium is then selectively isolated using ion exchange chromatography and the radioactivity is measured via liquid scintillation spectroscopy, with mass spectroscopy employed for yield quantification (see Fig. \ref{fig:flowchart}). The protocol's accuracy and reliability are validated through comparisons with established methods reported in the literature.

\section{Materials and Methods}

\subsection{Reagents and Apparatus}
All the reagents used in this study (HCl, TCA, NaOH, Na$_{2}$CO$_{3}$, H$_{2}$O$_{2}$, HNO$_{3}$, C$_{2}$H$_{2}$O$_{4}$) as well as the natural Sr tracer (ROTI\textsuperscript{\textregistered}Star, 1000 mg/L) used for yield measurements, were purchased from Carl Roth AG. Sr Resin\textsuperscript{\textregistered} chromatography columns (2~mL) were obtained from Eichrom Technologies Inc, while the$^{90}$Sr standard solution (5~mL, 1.85~MBq at 01.02.2022) was sourced from Eckert \& Ziegler Nuclitec GmbH. UHT full-fat cow milk (3.5$\%$ fat) was selected for this study to evaluate the protocol under conservative and challenging conditions, as the higher fat content increases the complexity of sample preparation. In contrast, using skimmed or semi-skimmed milk would likely simplify both calcination and purification steps, due to lower organic load.

Experimental setups included an Alpha 1-2 LDplus freeze dryer (Martin Christ GmbH) and a muffle furnace (Solo Switzerland) for calcination. Liquid scintillation measurements were performed using a TriCarb 2900TR spectrometer with Ultima Gold AB scintillation liquid (PerkinElmer). Chemical yields were determined via an Agilent 7700X ICP Mass Spectrometer.

\subsection{Freeze drying}
\begin{figure}[t!]
    \centering
    \includegraphics[width=.9\linewidth]{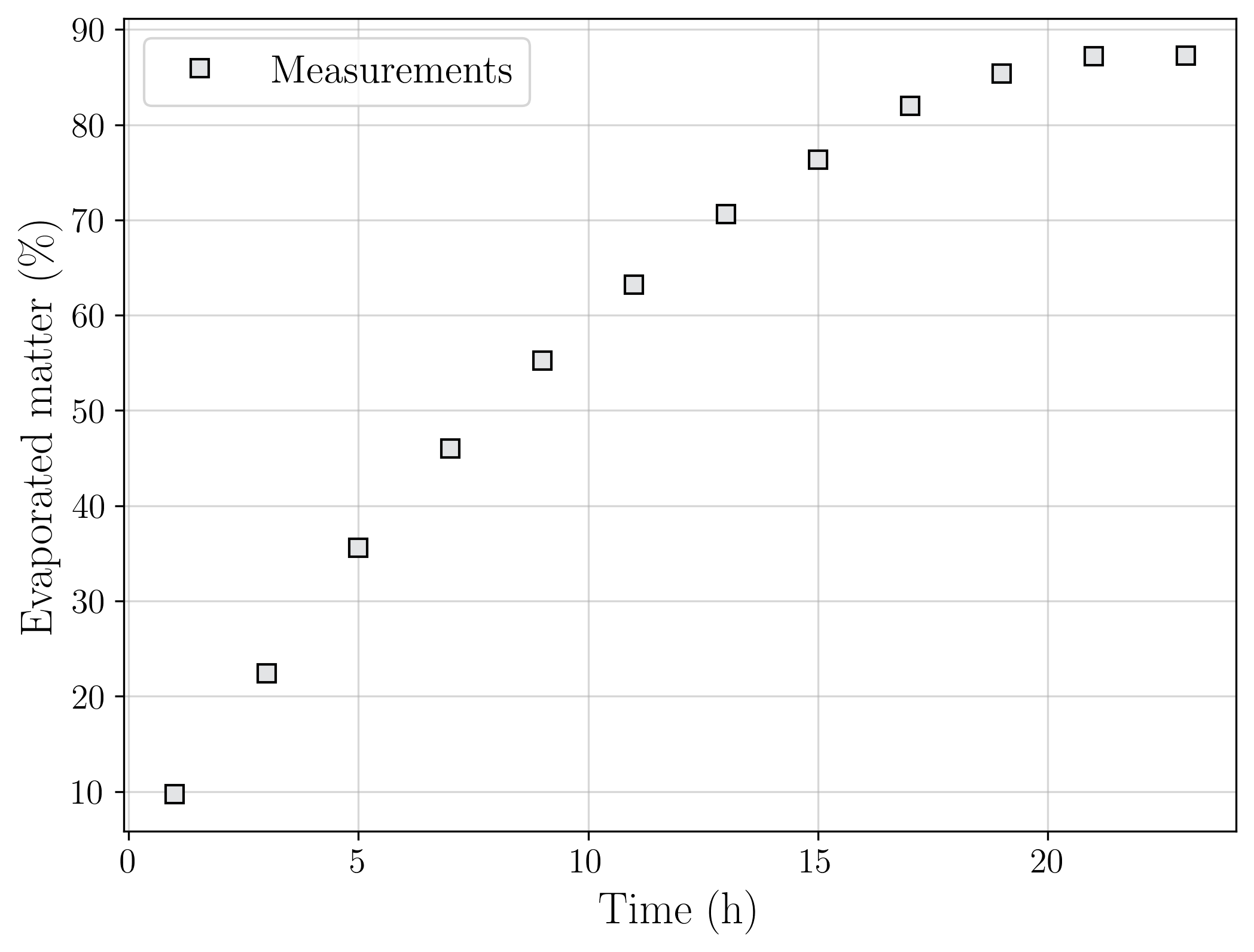}
	\caption{Evaporated matter as a function of time. After 20 h of freeze-drying, milk evaporation reaches a plateau at approximately 87$\%$. Incomplete evaporation results in the formation of a fragile dried milk block that adheres to the container walls and becomes volatile during the calcination process.}
	\label{fig:evaporation}
\end{figure}

Freeze-drying or else lyophilization, is the initial step of the developed protocol, used to convert liquid milk into dry powder and prevent issues associated with liquid milk calcination. The freeze-dryer used in this study is equipped with a three-tiered support system, allowing simultaneous processing of three large-diameter Pyrex crystallizers placed under a glass dome. The wide surface area minimizes sample thickness, thereby reducing drying time. Prior to freeze-drying, the milk samples are pre-frozen in a freezer overnight (approximately 18 hours).
The drying progress is monitored by visually inspecting the bottom of the crystallizer for the presence of residual liquid or ice. Upon completion, the milk forms a dry, easily detachable block, yielding approximately 12 g of powdered milk per 100 mL of liquid sample. The percentage of evaporated matter over time for a 100 mL crystallizer is illustrated in the Fig. \ref{fig:evaporation}.

Extensive testing was performed to maximize the evaporated matter by optimizing the freeze-drying parameters, specifically pre-freezing and drying temperatures, while maintaining a consistent sample volume (100~mL per crystallizer over 24~hours). Results indicated that sample position significantly affects drying efficiency, with the sample closest to the condenser (see sample 3 in Table \ref{table:lyophilisation}) exhibiting lower evaporation rates due to inadequate heat transfer during sublimation. The impact of pre-freezing and freeze-drying temperatures on the evaporated matter of the milk samples was found to be negligible, except for the sample positioned closest to the condenser. Based on these findings, the optimal configuration was determined to be pre-freezing at -20°C and freeze-drying at -40°C, ensuring efficient water removal while preserving the integrity of the powdered milk for subsequent processing steps.

\begin{table}[t!]
    \centering
    \begin{tabular}{ccccc}
    \toprule
    & \multicolumn{4}{c}{Evaporated matter (\%)} \\
    \midrule
        Pre-freezing & -40°C & -40°C & -20°C & -20°C \\
        Freeze drying & -40°C & -50°C & -40°C & -30°C \\
        \midrule
        Sample 1 & 87.29\% & 86.85\% & 87.33\% & 87.27\% \\ 
        Sample 2 & 86.58\% & 84.76\% & 87.10\% & 87.17\% \\ 
        Sample 3 & 83.98\% & 81.63\% & 86.29\% & 83.64\% \\ 
    \bottomrule
    \end{tabular}
    \caption{Percentage of evaporated matter as a function of pre-freezing and freeze-drying temperatures. Maximum evaporation is observed for pre-freezing at -20°C and freeze-drying at -40°C. The sample closer to the condenser exhibits lower evaporation rates due to inadequate heat transfer during sublimation.}
    \label{table:lyophilisation}
\end{table}

\subsection{Calcination}
\begin{figure}[h]
    \centering
    \includegraphics[width=.9\linewidth]{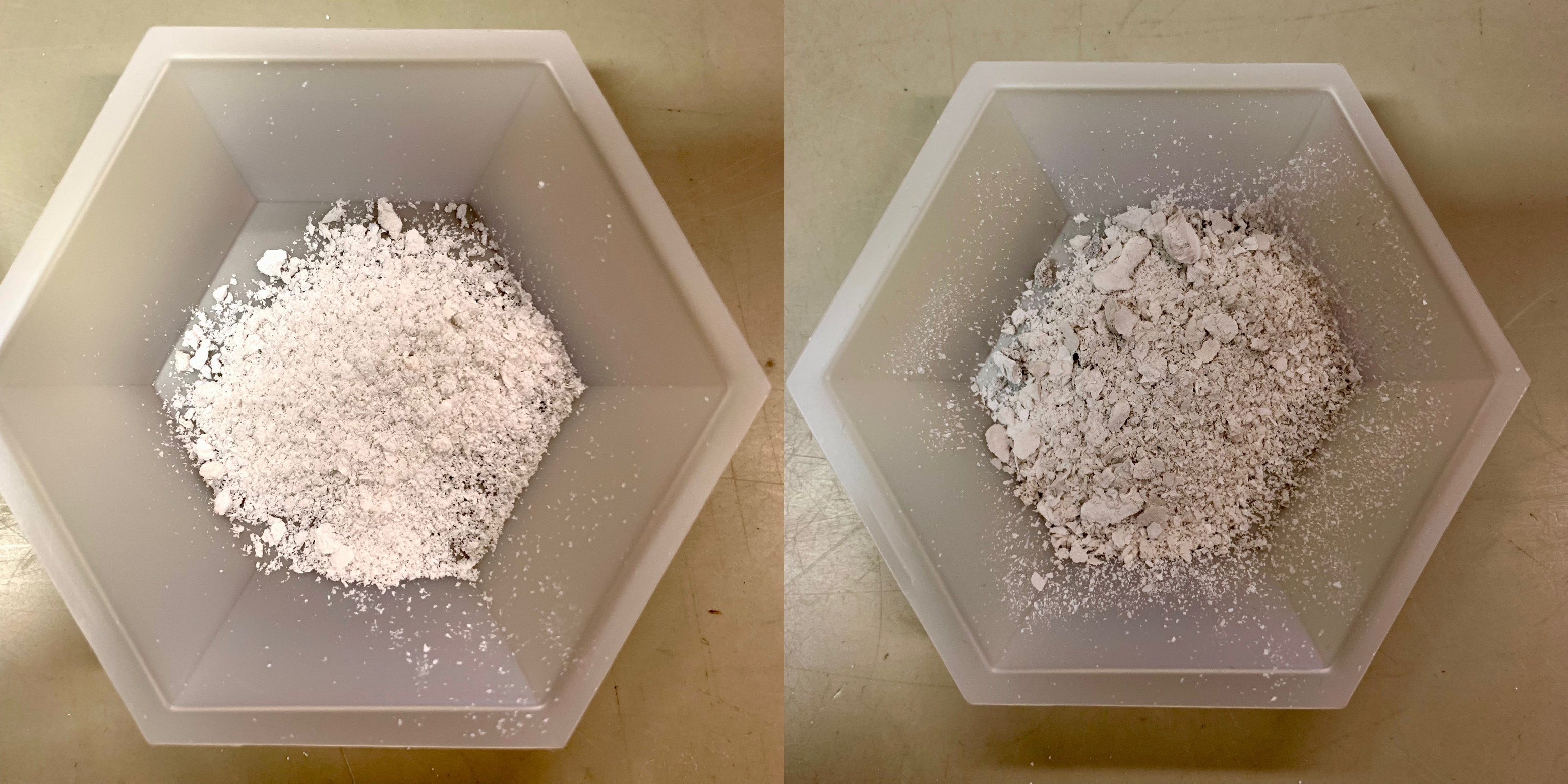}
	\caption{Colour of ashes after calcination in 700°C (left) and at 550°C (right). The brightness of the white colour indicates the extent of organic matter removal.}
	\label{fig:calcination}
\end{figure}

The powdered milk was transferred into a porcelain beaker and heated in a muffle for approximately 18~hours. Initially, calcination was performed at 550°C. However, organic matter removal was incomplete, yielding grayish ash (see Fig. \ref{fig:calcination}). The colour of the ash serves as a reliable indicator of residual organic materials, with bright white signifying complete organic matter removal. Furthermore, calcined ash obtained at 550°C clogged the resin column or significantly reduced the flow rate.

To address this issue, the calcination temperature was increased to 700°C. At this higher temperature, the resulting ash was white, indicating near-complete removal of organic matter. This optimization minimized interference during ion-exchange chromatography and ensured optimal separation efficiency. Furthermore, the flow rate during column separation aligned with the manufacturer’s specifications (0.6–0.8~mL/min), confirming the effectiveness of this temperature for reproducibility and yield.

For 12~gr of powdered milk (equivalent to 100~mL of liquid milk), approximately 0.6-0.7~g of ash was obtained, primarily consisting of natural mineral salts (potassium, calcium, and magnesium) present in milk, along with trace amounts of resistant organic matter \cite{bylund2015}.

\subsection{Ion Exchange chromatography}
Column chromatography was performed using a Sr Resin\textsuperscript{\textregistered} column (2~mL). The columns were mounted on a support, with beakers placed underneath to collect rinsing waste. According to the manufacturer (Triskem), the columns have a theoretical efficiency of 100\% for calcium loads up to 300~mg per 2~mL column, which corresponds to a maximum of 250~mL of milk per column (assuming a calcium content of 120~mg/100~mL of milk \cite{calcium}). 

The process begins with preconditioning the column using the same reagent as the sample solution. Since the calcined ash is diluted in 10~mL of 8M~HNO$_{3}$, the column is preconditioned by adding 5~mL of 8M~HNO$_{3}$.The diluted sample is then passed through the column. A series of rinses follows to separate strontium (stable or radioactive) from most other elements, based on their retention behaviour \cite{kfactor}, which varies with nitric acid concentration. The rinsing sequence begins with 15~mL of 8M~HNO$_{3}$, followed by 5~mL of 3M~HNO$_{3}$ mixed with 0.05M oxalic acid to eliminate potential actinides. This step is crucial, as oxalic acid facilitates the formation of soluble actinide complexes, preventing their retention on the Sr Resin\textsuperscript{\textregistered} column \cite{maxwell}. Finally, 7~mL of 8M~HNO$_{3}$ is used to complete the rinsing process.

The chromatography concludes with the elution stage, where strontium retained on the resin is recovered. This is achieved by using 10~mL solution of 0.05M~HNO$_{3}$, which is collected directly into a 20~mL liquid scintillation vial for subsequent measurements.

%\begin{figure}[t!]
%    \centering
	%\includegraphics[width=.9\linewidth]{facteurk.PNG}
	%\caption{Retention factor $k'$ of elements in Sr resin as a function of HNO$_3$ concentration \cite{kfactor}. At high HNO$_{3}$ concentrations, strontium is retained, while other metals pass through the column. Retained strontium can be eluted using low-concentration HNO$_{3}$ solutions.}
	%\label{fig:kfactor}
%\end{figure}

\subsection{Liquid Scintillation Spectroscopy}
\begin{figure}[t!]
    \centering
	\includegraphics[width=\linewidth]{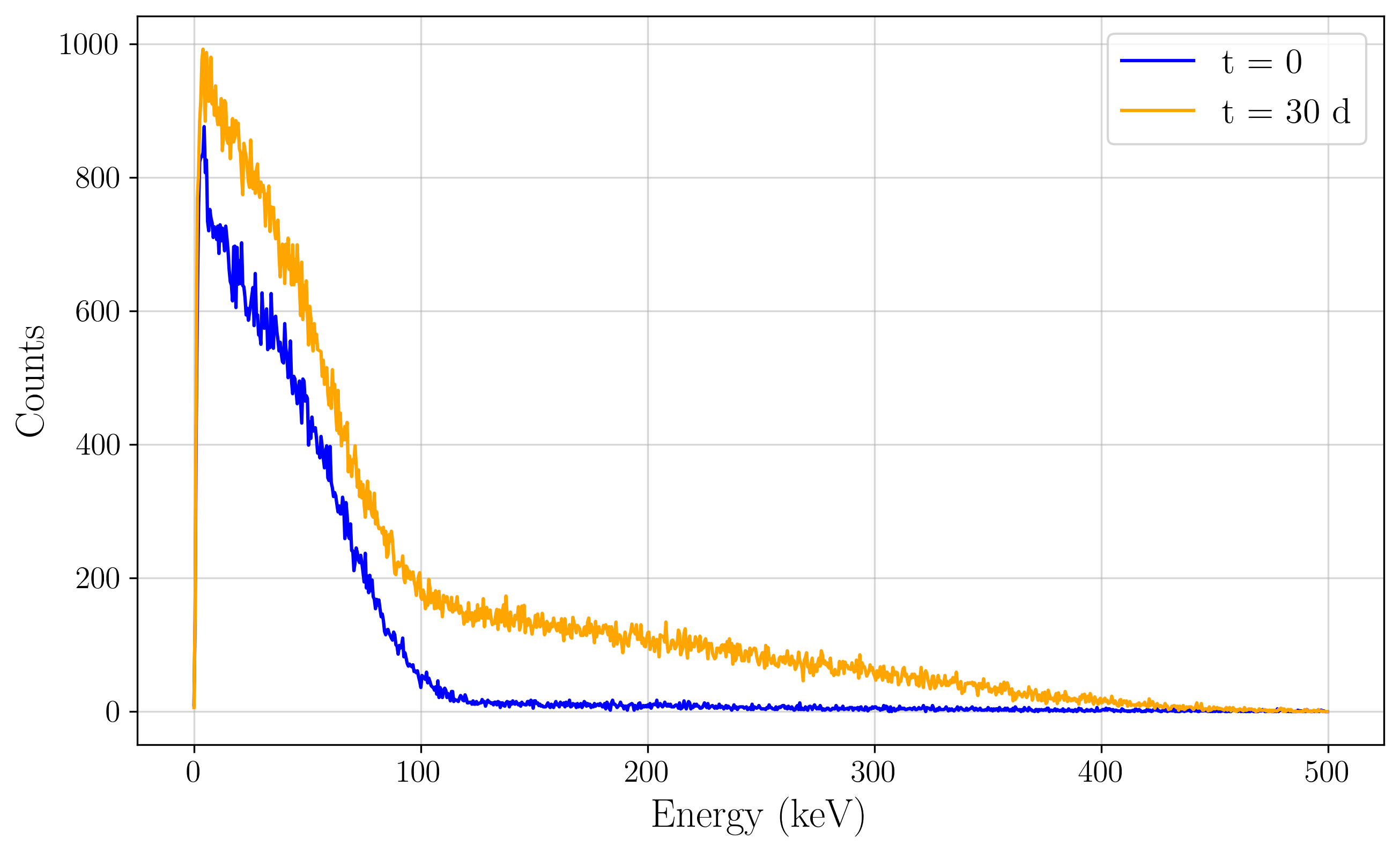}
    \caption{Liquid scintillation spectra of a $^{90}$Sr-contaminated milk sample. The blue curve represents the spectrum immediately after chromatography (t=0), where only $^{90}$Sr is present in the low energy region of the spectrum ($Q_{\beta^{-}}$ ($^{90}$Sr) = 545.9~keV). The orange curve represents the spectrum at t=30~days, after $^{90}$Sr and $^{90}$Y have reached secular equilibrium. At this point, $^{90}$Y appears in the high-energy region of the spectrum ($Q_{\beta^{-}}$ ($^{90}$Y) = 2278.5~keV).  \cite{Basu2020}}
	\label{fig:lsc_spectra}
\end{figure}

Liquid scintillation measurements were performed using a Packard TriCarb 2900TR spectrometer. The radioactive liquid eluted from the chromatography column was mixed with Ultima Gold AB scintillation liquid inside a 20 mL borosilicate glass vial. Samples were measured immediately after the chromatography process (t=0), twenty days later (t=20~d) to ensure that the $^{90}$Sr/$^{90}$Y pair reached secular equilibrium, and finally after 30 days (t=30~d). This approach allowed for a clearer distinction of each radioisotope's contribution to the final spectrum shape, thereby optimizing the final activity calculation (see Fig. \ref{fig:lsc_spectra}).

\subsection{Mass Spectrometry}

Mass spectrometry measurements were conducted using an Agilent 7700X instrument, paired with an ASX-500 auto-sampler. The liquid sample for analysis was contained in a 10~mL tube placed in the auto-sampler rack. Each measurement required an internal standard for accurate calibration, which was dispensed by the auto-sampler for each sample. The measurement process began with several system rinses using a mildly acidic solution (2\% v/v HNO$_{3}$). Following the rinsing procedure, samples were analysed based on their designated positions recorded in the software.

\begin{figure}[t!]
    \centering
	\includegraphics[width=\linewidth]{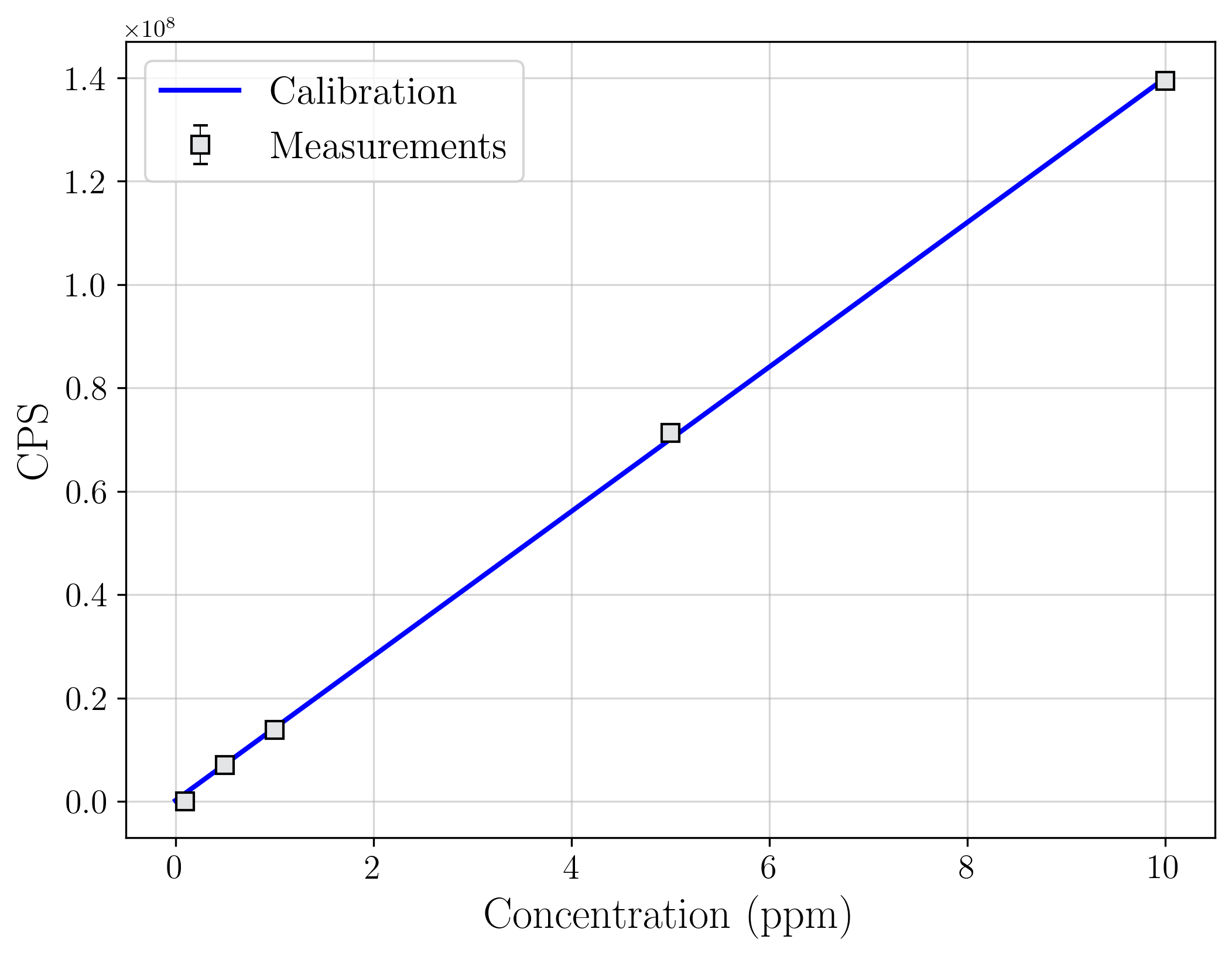}
	\caption{ICP-MS calibration curve. Five samples of aqueous solutions spiked with stable strontium tracer at various concentrations were prepared and measured.}
	\label{fig:icpms}
\end{figure}

The process consisted of two main steps. The first step involved calibrating the instrument using five aqueous solutions spiked with a stable strontium tracer at various concentrations. These samples were prepared and measured to generate a calibration curve for strontium concentrations ranging from 0.1 to 10~ppm (see Fig. \ref{fig:icpms}).

In the second step, milk samples were spiked with 1~mg of stable strontium tracer per 100~mL of milk. The protocol described in this study was then applied, with ICP-MS measurements performed before and after each step of organic matter removal and strontium separation. This approach enabled the calculation of strontium recovery ratio relative to its initial concentration. Since $^{90}$Sr exhibits chemical behaviour similar to its stable isotopes it was assumed that the recovery percentage of strontium remains consistent across all isotopes. This assumption allowed the recovered stable strontium ratio in milk to be extrapolated to $^{90}$Sr, facilitating protocol efficiency assessment and adjustment of activity values measured by liquid scintillation spectroscopy.

Moreover, ICP-MS enables the detection of trace amounts of other elements within the sample. This capability was used to assess whether other elements could interfere with the activity values obtained from liquid scintillation spectroscopy.

% \begin{table}[t! ]
%     \centering
%     \begin{tabular}{cc}
%     \toprule
%         \textbf{Element}  & \textbf{Decontamination factor}  \\ 
%         \midrule
%         Na & 617 ± 99 \\ 
%         Mg & 2721 ± 668 \\ 
%         K & 1369 ± 232 \\
%         Ca & 16 ± 4 \\ 
%         Ba & 46 ± 2 \\ 
%         Pb & 159 ± 59 \\ 
%         Th & 25 ± 8 \\ 
%         Cl & 46 ± 5 \\ 
%         Zr & 10 ± 1 \\ 
%         Mo & 106 ± 20 \\ 
%     \bottomrule
%     \end{tabular}
%     \caption{Decontamination factor of the main elements after passing through the Sr Resin\textsuperscript{\textregistered} column.}
% 	\label{fig:DF}
% \end{table}

\section{Results and Discussion}
As part of an intercomparison measurement between several certified Swiss nuclear laboratories, a milk sample containing $^{90}$Sr at an unknown concentration was received, known to be below the Swiss release limit for this radionuclide (1 Bq/mL, as defined in the Swiss Radiological Protection Ordinance, ORaP \cite{orap}). To replicate a similar $^{90}$Sr concentration to that expected in the received samples, reference milk samples were prepared with a concentration of 0.45~Bq/mL. The protocol was then applied, achieving a high efficiency of 100$\pm$2\%  as measured by ICP-MS, demonstrating the method’s reproducibility and robustness. These results confirm that the developed protocol retains 100\% of the added $^{90}$Sr, enabling accurate quantification through liquid scintillation spectroscopy.

It is important to note that ICP-MS measurements was performed solely to validate the chemical yield of the protocol at 100\%. Routine mass spectrometry is not required when applying this protocol, as findings indicate that the chemical yield consistently achieves near-complete recovery.

\begin{figure}[t!]
    \centering
    \includegraphics[width=\linewidth]{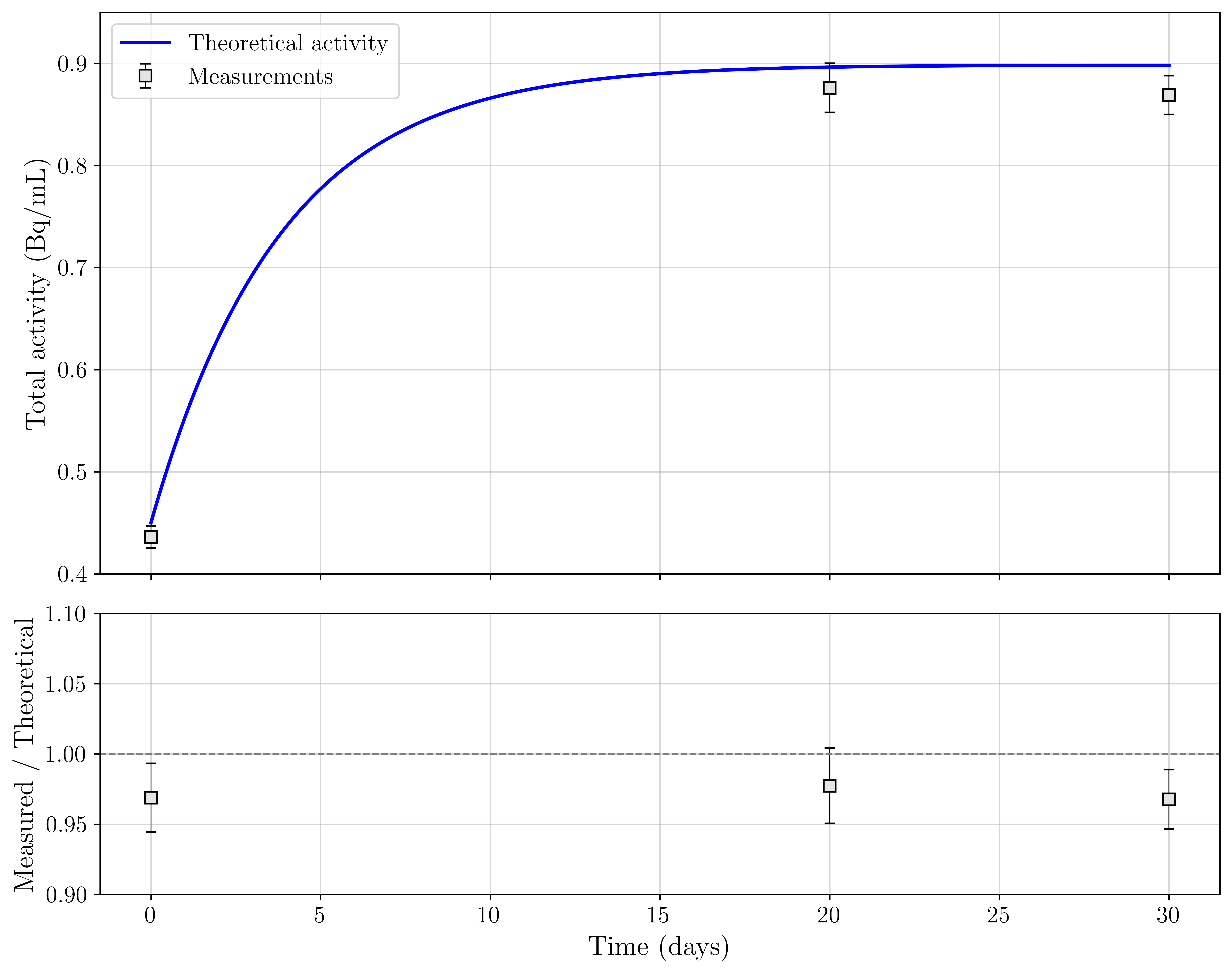}
    \caption{Total activity evolution of radiocontaminated milk samples with $^{90}$Sr over 30 days (top) and ratio of measured to theoretical total activity (bottom) following chemical separation of $^{90}$Sr from milk samples. Secular equilibrium with its daughter $^{90}$Y ($T_{1/2} = 2.67$~d) is expected after approximately 14 days, resulting in a doubling of total activity. The parent radionuclide $^{90}$Sr ($T_{1/2} = 28.8$~y) remains constant over this period. The agreement between measured and expected activity growth confirms effective separation of $^{90}$Sr and $^{90}$Y during column chromatography, and the absence of other significant beta emitters.}
    \label{fig:growth}
\end{figure}

To further validate the protocol’s accuracy, more LSC measurements were conducted at 20 and 30 days to track the increase in $^{90}$Y activity until secular equilibrium was reached. At this point, the absolute activity of all measured samples doubled and then stabilized, confirming the effective separation of strontium and yttrium during chromatography (see Fig. \ref{fig:growth}).

The MDA (Minimum Detectable Activity) depends on three factors: the amount of milk used in the protocol ($V$), the efficiency of the protocol ($\eta$), and the detection limit of the measurement device ($A_{LSC}$), which itself depends on the measurement time.

\begin{equation*}
\text{MDA} = \frac{A_{{LSC}} \cdot \eta}{V}
\end{equation*}

Assuming our protocol is applied with $V$ = 300~mL of milk, $\eta$ $\approx$ 100\% efficiency and $A_{LSC}$ = 0.1 Bq for a 60-minute measurement, the calculated MDA is 0.33~Bq/L. This low MDA demonstrates the protocol’s applicability for detecting low-level $^{90}$Sr contamination, which is crucial for effective environmental monitoring. This value could be significantly reduced using modern Tricarb spectrometers equipped with 'Ultra Low Level' spectroscopy capabilities, which typically detect count rates as low as 1–20 CPM above background. Such advancements make the protocol highly suitable for rapid-response scenarios, requiring precise detection at low contamination levels.

\begin{figure}[t!]
    \centering
    \includegraphics[width=\linewidth]{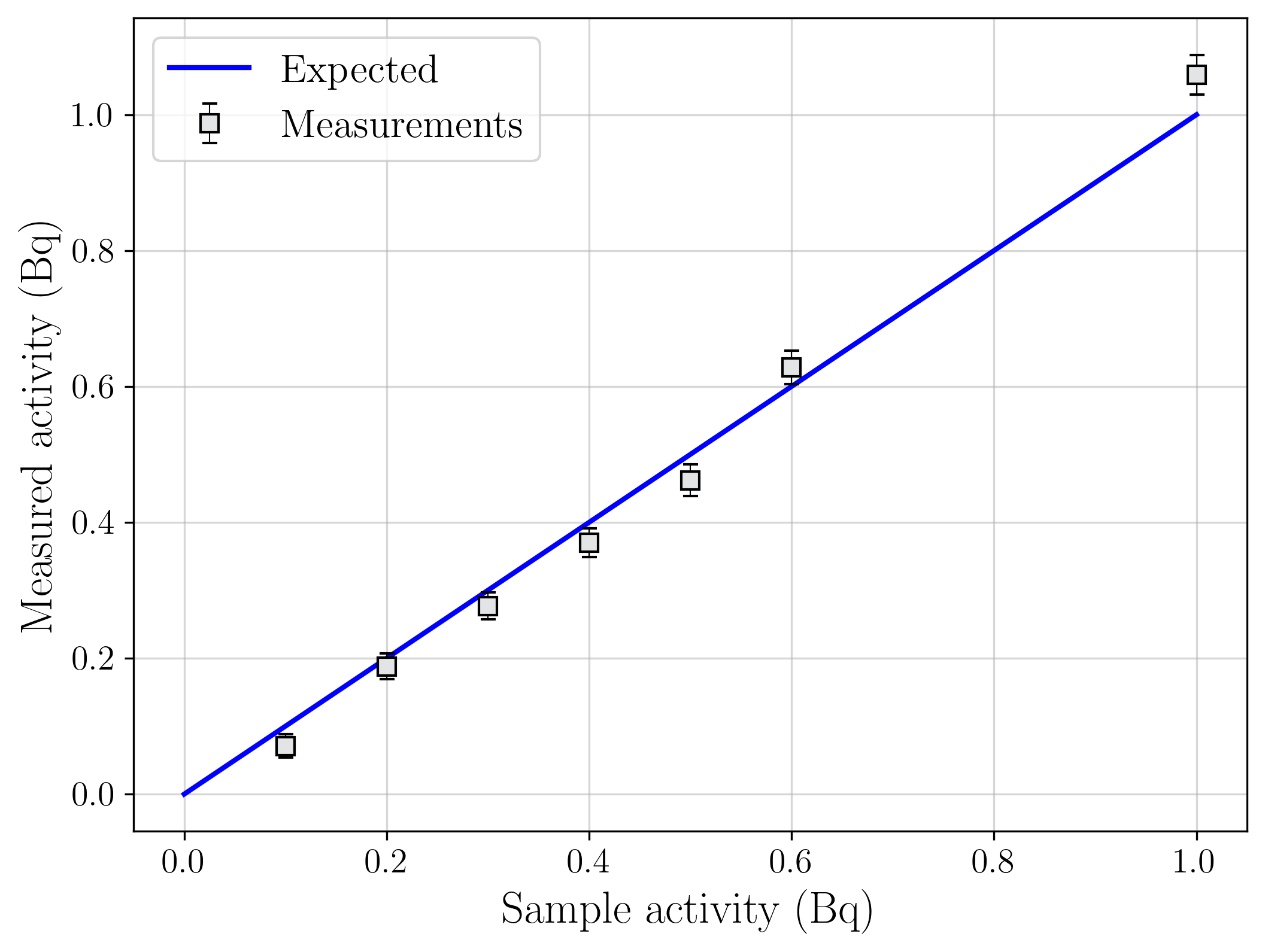}
    \caption{Comparison of measured activity versus reference activity for $^{90}$Sr-spiked aqueous samples using liquid scintillation spectroscopy. The graph demonstrates the accuracy and linearity of the measurement protocol, ensuring consistent and precise values down to 0.1~Bq.}
    \label{fig:MDA}
\end{figure}

The detection limit of 0.1~Bq was determined by measuring a series of $^{90}$Sr-spiked aqueous samples, ranging from 0.1 to 1~Bq (see Fig.~\ref{fig:MDA}), using the liquid scintillation spectrometer. These water-based standards were used to simulate the final eluates obtained after column separation, in order to assess the instrumental detection limit independently of the milk matrix. This approach allows for the calculation of the MDA under realistic laboratory conditions and highlights the role of detector sensitivity in determining overall method performance.

\subsection{Comparison to other methods}
Methods for separating strontium and measuring its activity are generally similar in the literature. Strontium is typically isolated using column chromatography, such as Sr Resin\textsuperscript{\textregistered}, followed by activity measurement via liquid scintillation spectroscopy, Cherenkov counting, or $\beta$ counting. The main variation between different protocols lies in organic matter removal process. We tested three methods for organic matter removal: two centrifugation based techniques \cite{Guerin2017,maxwell}, and one method using Dowex\textsuperscript{\textregistered}50W-X8 cation exchange resin, which is widely used in radiochemistry for food sample preparation \cite{dowex1,dowex2}.  

\begin{table}[t!]
    \centering
    \begin{tabular}{cc}
    \toprule
        Method & Sr Yield (\%) \\ 
        \midrule
        Guérin & 68 ± 2 \\
        Maxwell & N/A \\
        Dowex\textsuperscript{\textregistered} 50W-X8 & 65 ± 6 \\
        Present study & 100 ± 2 \\
    \bottomrule
    \end{tabular}
    \caption{Comparison of strontium yields from different tested methods.}
    \label{fig:comparison}
\end{table}

\subsection{The Guérin protocol} 
This method involves centrifugation techniques combined with novel chemical reagents combinations. First, milk coagulation is induced by mixing 12M~HCl and 50\%~w/v~TCA, with the supernatant retained after centrifugation. In the second step, strontium is precipitated as carbonate (SrCO$_{3}$) by adding 15\%~w/v~Na$_{2}$CO$_{3}$ after adjusting the solution to pH = 12 using  40\%~w/v~NaOH. The precipitate is then retained after centrifugation and dissolved in 8M~HNO$_{3}$ for chromatographic separation, a step common to all protocols. An ICP-MS yield of 68$\pm$2\% was measured from seven samples, consistent with results of Guérin et al. \cite{Guerin2017}.

\subsection{The Maxwell protocol} 
This second method also involves centrifugation techniques but uses different chemical reagents. Unlike the Guérin protocol, the Maxwell method begins with strontium precipitation before organic matter removal. To achieve this, 1.25M~Ca(NO$_{3}$)$_{2}$ and 3.2M~(NH$_{4}$)$_{2}$HPO$_{4}$ are added, followed by pH adjustment to 10 using gradual addition of NH$_{4}$OH. The sample is then centrifuged, and the precipitate containing strontium is retained. Subsequently, 3M~HNO$_{3}$ is added to dissolve the precipitate, coagulating the fats and proteins after centrifugation.

At this stage, an issue arose, the precipitate was less dense than the supernatant, making it difficult to recover the liquid containing strontium. As a result, a significant portion of organic matter remained in the liquid. The protocol continues with evaporating the supernatant, performing a first wet calcination, followed by calcination at 550°C and a second wet calcination. Wet calcination involves mixing 15.7M~HNO$_{3}$ with 30\%~w/t~H$_{2}$O$_{2}$, then evaporating the mixture on a hot plate to eliminate organic residues. However, when using the reagent quantities recommended by Maxwell et al. \cite{maxwell}, wet calcination proved time-consuming, and relatively hazardous, both in terms of safety and radiological protection. 

\subsection{Dowex\textsuperscript{\textregistered}50W-X8 Method}
The third method investigated employs Dowex\textsuperscript{\textregistered} 50W-X8 cation exchange resin, following a protocol based on conventional techniques previously reported in the literature \cite{dowex1,dowex2}. This method is designed to effectively separate strontium while minimizing organic interference. The procedure consists of the sample preparation using 500~mL of milk, 60~g of resin, and 2~mg of natural strontium ($^{nat}$Sr) that are mixed in a beaker and left to rest for 1~hour to allow ion exchange. The supernatant is then discarded, and 100~mL of 0.01M~HNO$_{3}$ is added and mixed for 5 minutes. The mixture is left to rest for 30 minutes, after which the supernatant is removed. A second 100~mL of 0.01M~HNO$_{3}$ is added and then transferred into a chromatography column.

To elute the strontium retained on the resin, 150~mL of 8M~HNO$_{3}$ is passed through the column. The solution is collected and evaporated to dryness. If organic matter removal is complete, the residue appears whitish, indicating that only the naturally occurring minerals from milk remain. The remaining residue is dissolved in 8M~HNO$_{3}$ for subsequent chromatographic separation and analysis. An ICP-MS yield of 65$\pm$6\% was measured across six samples, validating the effectiveness and reproducibility of this method for strontium separation and quantification.

\section{Conclusions and Outlook}

This study presents a robust and efficient protocol for the quantification of $^{90}$Sr in milk, addressing critical food safety concerns related to potential radioactive contamination. The proposed method integrates freeze-drying, high-temperature calcination (at 700\textdegree{}C), ion exchange chromatography using Sr Resin\textsuperscript{\textregistered}, and liquid scintillation spectroscopy. It achieves a high recovery efficiency of 100$\pm$2\%, as determined by ICP-MS measurements. Theoretical calculations indicate a Minimum Detectable Activity (MDA) of 0.33~Bq/L under optimal conditions. Additionally, the protocol was experimentally validated at a concentration of 0.45~Bq/mL, demonstrating its practical suitability for detecting low-level $^{90}$Sr contamination in milk, particularly during nuclear emergencies.

Beyond its performance metrics, the strength of this protocol lies in the reliability and reproducibility of its core steps, which include freeze-drying, calcination, and strontium-specific separation. These stages are essential for the elimination of organic matter and the selective extraction of strontium, and they can be applied to a broad range of radionuclides in food matrices. Unlike chemical digestion or precipitation techniques, this protocol relies mainly on physical processing steps and minimal reagent use, making it safer to implement, less operator-dependent, and more cost-effective. Furthermore, Sr Resin\textsuperscript{\textregistered} columns can be reused multiple times without significant loss of performance, further reducing operational costs \cite{dowex2,jakopivc2005tracer,de2009systematic}.

Comparative testing with other methods such as the Guérin and Dowex\textsuperscript{\textregistered} 50W-X8 protocols highlighted the challenges of achieving high and consistent yields while ensuring safe handling conditions. The proposed method was developed with a focus on reproducibility, safety, and ease of use, especially when dealing with radioactive samples. While other methods may offer shorter processing times, they often require intensive manual handling or multiple centrifugation steps, which can limit scalability and introduce variability.

Although designed with nuclear emergency contexts in mind, this protocol is especially well suited for routine monitoring of large sample batches. Freeze-drying and calcination are easily scalable and compatible with high-throughput workflows, unlike centrifugation-based methods, which require repeated manual interventions to process larger volumes. Additionally, while calcination may appear time-consuming, the process involves primarily equipment time, not operation time, reducing workload and exposure.

Future work could adapt this protocol to other food matrices such as vegetables, grains, or processed products. Pairing it with more advanced detection systems such as ultra-low-level liquid scintillation counters would further improve sensitivity and extend its applicability to very low contamination levels. Overall, this method offers a practical, safe, and scalable framework for reliable $^{90}$Sr monitoring in food, contributing to improved preparedness and public health protection in radiological contexts.

\section*{Acknowledgements}
The authors would like to express their sincere gratitude to the inTECH institute of the School of Landscape, Architecture, and Engineering (HEPIA), of University of Applied Sciences and Arts Western Switzerland for their invaluable support throughout this research. The provision of equipment and laboratory time to our nuclear physics and nuclear chemistry laboratories was instrumental in the successful completion of this study.

% \appendix
% \section{Example Appendix Section}


% %% For citations use: 
% %%       \cite{<label>} ==> [1]

% %%
% Example citation, See \cite{lamport94}.




\bibliographystyle{elsarticle-num} 
\bibliography{Sr90_milk_anp2024}

% %% For numbered reference style

%   \textit{\LaTeX: a document preparation system},
%   Addison Wesley, Massachusetts,

% \end{thebibliography}
\end{document}